\documentstyle{mn}

\newif\ifAMStwofonts

\ifoldfss
  \ifCUPmtlplainloaded \else
    \NewTextAlphabet{textbfit} {cmbxti10} {}
    \NewTextAlphabet{textbfss} {cmssbx10} {}
    \NewMathAlphabet{mathbfit} {cmbxti10} {} 
    \NewMathAlphabet{mathbfss} {cmssbx10} {} 
  \fi
  \ifAMStwofonts
    \ifCUPmtlplainloaded \else
      \NewSymbolFont{upmath} {eurm10}
      \NewSymbolFont{AMSa} {msam10}
      \NewMathSymbol{\upi}     {0}{upmath}{19}
      \NewMathSymbol{\umu}     {0}{upmath}{16}
      \NewMathSymbol{\upartial}{0}{upmath}{40}
      \NewMathSymbol{\leqslant}{3}{AMSa}{36}
      \NewMathSymbol{\geqslant}{3}{AMSa}{3E}

      \let\leq=\leqslant 
      \let\geq=\geqslant 
    \fi
  \fi
\fi 

\ifnfssone
  \newmathalphabet{\mathit}
  \addtoversion{normal}{\mathit}{cmr}{m}{it}
  \addtoversion{bold}{\mathit}{cmr}{bx}{it}
  \newmathalphabet{\mathbfit} 
  \addtoversion{normal}{\mathbfit}{cmr}{bx}{it}
  \addtoversion{bold}{\mathbfit}{cmr}{bx}{it}
  \newmathalphabet{\mathbfss} 
  \addtoversion{normal}{\mathbfss}{cmss}{bx}{n}
  \addtoversion{bold}{\mathbfss}{cmss}{bx}{n}
  \ifAMStwofonts
    \ifCUPmtlplainloaded \else
      \UseAMStwoboldmath
      \makeatletter
      \new@mathgroup\upmath@group
      \define@mathgroup\mv@normal\upmath@group{eur}{m}{n}
      \define@mathgroup\mv@bold\upmath@group{eur}{b}{n}
      \edef\UPM{\hexnumber\upmath@group}
      \new@mathgroup\amsa@group
      \define@mathgroup\mv@normal\amsa@group{msa}{m}{n}
      \define@mathgroup\mv@bold\amsa@group{msa}{m}{n}
      \edef\AMSa{\hexnumber\amsa@group}
      \makeatother
      \mathchardef\upi="0\UPM19
      \mathchardef\umu="0\UPM16
      \mathchardef\upartial="0\UPM40
      \mathchardef\leqslant="3\AMSa36
      \mathchardef\geqslant="3\AMSa3E

      \let\leq=\leqslant 
      \let\geq=\geqslant 
    \fi
  \fi
\fi 

\ifnfsstwo
  \DeclareMathAlphabet{\mathbfit}{OT1}{cmr}{bx}{it}
  \SetMathAlphabet\mathbfit{bold}{OT1}{cmr}{bx}{it}
  \DeclareMathAlphabet{\mathbfss}{OT1}{cmss}{bx}{n}
  \SetMathAlphabet\mathbfss{bold}{OT1}{cmss}{bx}{n}
  \ifAMStwofonts
    \ifCUPmtlplainloaded \else
      \DeclareSymbolFont{UPM}{U}{eur}{m}{n}
      \SetSymbolFont{UPM}{bold}{U}{eur}{b}{n}
      \DeclareSymbolFont{AMSa}{U}{msa}{m}{n}
      \DeclareMathSymbol{\upi}{0}{UPM}{"19}
      \DeclareMathSymbol{\umu}{0}{UPM}{"16}
      \DeclareMathSymbol{\upartial}{0}{UPM}{"40}
      \DeclareMathSymbol{\leqslant}{3}{AMSa}{"36}
      \DeclareMathSymbol{\geqslant}{3}{AMSa}{"3E}

      \let\leq=\leqslant 
      \let\geq=\geqslant 
    \fi
  \fi
\fi 

\ifCUPmtlplainloaded \else
  \ifAMStwofonts \else 
    \def\upi{\pi}
    \def\umu{\mu}
    \def\upartial{\partial}
  \fi
\fi

\title{Population synthesis of H{\Large {\bf II}}  galaxies}
\author[D. Raimann et al.]
  {D.~Raimann$^1$, E.~Bica$^1$, T.~Storchi-Bergmann$^1$,  J.~Melnick$^2$
  \newauthor
  and H.~Schmitt$^3$ \thanks{E-mail: raimann@if.ufrgs.br; bica@if.ufrgs.br; thaisa@if.ufrgs.br; jmelnick@eso.org; schmitt@stsci.edu} \\
  $^1$ Universidade Federal do Rio Grande do Sul, IF, CP15051, Porto Alegre 91501-970, RS, Brazil \\
  $^2$ European Southern Observatory, Casilla 19001, Santiago 19, Chile \\
  $^3$ Space Telescope Institute, 3700 San Martin Drive, Baltimore, MD21218, USA} 

\date{Accepted  
      Received 
      in original form}

\pagerange{\pageref{firstpage}--\pageref{lastpage}}
\pubyear{1999}

\begin{document}

\maketitle

\title{Population synthesis of H{\normalsize {\it II}}  galaxies}

\label{firstpage}

\begin{abstract}
We study the stellar population of galaxies with active star formation, determining ages of the stellar components by means of spectral population synthesis of their absorption spectra. The data consist of optical spectra of 185 nearby ($z \leq 0.075$) emission line galaxies (Terlevich et al., 1991). They are mostly H{\sevensize II} galaxies, but we also include some Starbursts and Seyfert 2's, for comparison purposes. They were grouped into 19 high signal-to-noise ratio template spectra, according to their continuum distribution, absorption and emission line characteristics. The templates were then synthesized with a star cluster spectral base.

The synthesis results indicate that H{\sevensize II} galaxies are typically age-composite stellar systems, presenting important contribution from generations up to as old as 500 Myr. We detect a significant contribution of populations with ages older than 1 Gyr in two groups of H{\sevensize II} galaxies. The age distributions of stellar populations among Starbursts can vary considerably despite similarities in the emission line spectra. In the case of Seyfert 2 groups we obtain important contributions of old population, consistent with a bulge. 

From the diversity of star formation histories, we conclude that typical H{\sevensize II} galaxies in the local universe {\it are not} systems presently forming their first stellar generation.

\end{abstract}


\section{Introduction}

H{\sevensize II} galaxies are gas-rich dwarf galaxies undergoing violent star formation characterized by an important population of massive stars with strong narrow emission lines (Melnick et al., 1985, for a review see Melnick, 1992). Their study began with the discovery by Sargent and Searle (1970) that some compact galaxies in Zwicky's Catalogue (1971) had spectra comparable to those of giant H{\sevensize II} regions. 

Many of these objects have a compact morphology with a dominant starbursting region, but some have double or multiple components (Telles et al., 1997). Terlevich et al. (1991) built a catalogue of H{\sevensize II} galaxy spectra and analysed the properties of the individual objects, such as distributions of redshift, Balmer decrement, H$\beta$ luminosity and equivalent width. In particular, they derived $-1.4<$[O/H]$<-0.4$, showing that H{\sevensize II} galaxies are typically metal deficient. 

An important question is whether H{\sevensize II} galaxies are young systems that are forming their first generation of stars, or systems that have multiple stellar generations, including several Gyr old ones. Several studies have searched for an underlying old stellar component but the results were not conclusive (Thuan, 1983; Campbell and Terlevich, 1984; Melnick et al., 1985; Farnell et al., 1988). It would be also worth probing whether non-ionizing blue stellar populations (20--500 Myr) occur and in which proportions.

Brinks and Klein (1988) found that one H{\sevensize II} galaxy (II Zw 40) may be indeed a young system, recently formed by interaction of two H{\sevensize I} clouds. On the other hand, Schulte-Ladbeck and Crone (1998) resolved red giant stars in the blue compact dwarf galaxy VII Zw 403 with HST, and concluded that it is not a young galaxy forming its first generation of stars, having an older underlying component. Indeed such red giants could be related to an intermediate or old age component since the colour-magnitude diagram could not distinguish such low turnoffs.  

The spectral study of stellar populations requires a high enough signal-to-noise (S/N) ratio to clearly establish the continuum distribution and absorption lines. Averaging spectra of similar properties has proven to be a powerful tool for stellar population analyses of star clusters and galaxies (e.g. Bica, 1988; Bonatto et al., 1995; Santos et al., 1995).

The population synthesis of galaxies using a base of star cluster spectra (Bica, 1988) has provided information on population components in normal galaxies. The method employs a grid of star cluster spectral properties built as a function of age and metallicity (Bica and Alloin, 1986a, 1986b). The synthesis provides flux fractions for the successive generations of star formation and corresponding chemical enrichment. The method was also applied to the dusty metal-rich Starburst galaxy NGC6240 (Schmitt et al., 1996). Recently the algorithm was adapted to the Ultraviolet range (Bonatto et al., 1998, 1999).

In this Paper we use the star cluster synthesis method to investigate the stellar populations of H{\sevensize II} galaxies and to compare them with those of more luminous emission-line galaxies such as Starbursts and Seyfert 2's. We intend to shed light on the question about H{\sevensize II} galaxies as single or multigeneration systems. We present the sample and spectral group construction in Section 2. The population syntheses are carried out and discussed in Section 3. The concluding remarks are given in Section 4. We defer consideration of the emission line spectra to a forthcoming paper (Raimann et al., 1999, hereafter Paper II).

\section{Data}

The data set consists of 185 emission line galaxy spectra in the range $\lambda\lambda3300-7000$ \AA, with average resolution $\approx 5$ \AA. Most spectra (156) are a sub-sample of those obtained by Terlevich et al. (1991) for the spectrophotometric catalogue of H{\sevensize II} galaxies and analyses of the emission properties of the individual objects. The present sample includes 29 additional objects, mostly Starbursts and Seyfert 2 galaxies, obtained in the same observing runs. The selected spectra are those with enough continuum S/N ratio ($\approx$ 5--10) to allow one to distinguish the basic spectral distribution of the stellar population.

Prior to any spectral averaging among galaxies, individual spectra were corrected for the foreground (Milky Way) reddening (Burstein and Heiles, 1984) using a normal reddening law (Seaton, 1979). Subsequently, they were redshift corrected with radial velocities from Terlevich et al. (1991), the NASA/IPAC Extragalactic Database (NED)\footnote{The NASA/IPEC Extragalactic Database (NED) is operated by the Jet Propulsion Laboratory, California Institute of Technology, under contract with the National Aeronautics and Space Administration.}, or derived from the spectra themselves. Finally, the individual spectra were normalized to F$_{\lambda}=10$ at $\lambda$5870 \AA.

Many spectra had similar characteristics, allowing one to classify them into groups. We follow the spectral grouping procedures already applied to nuclear regions of early and late type normal galaxies (Bica, 1988), and more recently to the UV range of early type galaxies (Bonatto et al., 1996),  nuclear starbursts of spiral galaxies (Bonatto et al., 1998) and star forming galaxies in general (Bonatto et al., 1999). In the present study we considered the continuum spectral distribution, absorption features whenever seen (especially the Balmer jump), and emission lines. In the latter case we measured equivalent widths (W$_\lambda$) of the strongest emission lines [O{\sevensize II}]$\lambda3727$, H$\beta$, [O{\sevensize III}]$\lambda4959$ (and assuming a ratio 1:3 we inferred the $\lambda5007$ line, thus avoiding any saturation effect), H$\alpha$ and [N{\sevensize II}]$\lambda6584$. In addition to spectral properties we adopted as grouping criteria comparable absolute magnitudes (M$_B$) and H$\beta$ emission line luminosities.

The criteria above, together with information on spectral (H{\sevensize II}, Starburst or Seyfert 2) and/or morphological (spiral, irregular, BCD, interacting, etc) characteristics, basically preclude that differently evolved galaxies be mixed in a given group. The sources for H$\beta$ fluxes, B integrated magnitudes, morphological and spectral types were Terlevich et al. (1991), the NED database and Sandage and Tammann (1981). The spectral averages were weighted according to the continuum S/N ratio of the individual spectra. Finally, we point out that each averaging stage among galaxy spectra improved the S/N ratio, and these partial results were checked for compatibility before further averaging.  

The resulting 19 spectral groups with their members are presented in Table \ref{memb}. By columns: (1) galaxy designation, (2) morphological and/or spectral types, (3) redshift, and (4) absolute magnitude (M$_B$) assuming $H_0 = 75$ km s$^{-1}$ Mpc$^{-1}$. Each group in the table is named after the member galaxy with the best S/N ratio, and we also present in the table basic group characteristics, namely emission line classification, average observed region (kpc) corresponding to the employed slit and respective galaxy redshift, and finally average absolute magnitude ($<$M$_B$$>$).    

We emphasize that the absolute magnitudes in Table \ref{memb} are indicative since the sources of B magnitudes are heterogeneous, varying from photographic estimates in surveys and detailed photographic analyses for bright galaxies to CCD imaging analyses. In some cases (e.g. NGC1487) the extensions in column 1 indicate clumps in the same object, so we adopted the absolute magnitude of the ensemble galaxy (column 4). Depending on the radial velocities within each group the basic spatial region observed varies from $0.2 \times 0.3$ kpc to $2.5 \times 3.8$ kpc with average value $1.1 \times 1.9$ kpc. 

Note that we are typically dealing with compact sources, in many cases they are nearly stellar in survey plates (e.g. the UM Catalogue, MacAlpine et al., 1977). Deep CCD imaging of H{\sevensize II} galaxies and structural analyses are available for relatively small samples (e.g. Telles and Terlevich, 1997; Telles et al., 1997). In particular Telles et al. (1997) classify H{\sevensize II} galaxies into Type I which are more disturbed extended having larger luminosities and velocity dispersions, and more compact and regular Type II objects. We also recall in column 2 this classification. Note that G\_Tol1004-296 has a significant number of galaxies in common with Telles et al. (1997) being typically of Type II, while G\_Cam1148-2020 has five galaxies in common, all of Type I. The morphological/spectral information in Table \ref{memb} is also interesting to see how often galaxies in each group have been classified as H{\sevensize II} (most groups), how in certain groups spiral morphology is mentioned  with peculiarities which could suggest the starburst phenomenon (e.g. G\_UM477), and finally other groups wherein Seyfert 2's are mentioned (e.g. G\_NGC4507).

\begin{table}
 \caption{The resulting spectral groups and their member galaxies.}
 \label{memb}
 \begin{tabular}{lrcc} \hline
Designation    & Morphology/   & z & M$_{B}$ \\
 & Spectral types & & \\ \hline
\multicolumn{4}{l}{{\bf G\_Cam1148-2020} H{\sevensize II}  $1.5 \times 1.8$ kpc  $<$M$_B$$>$$=-17.8 \pm 1.1$} \\ \hline
Cam1148-2020$^*$   &	 H{\sevensize II} TI		& 0.012 & \\
Cam0341-4045E$^*$  &	 H{\sevensize II} TI		& 0.015 & \\
Tol 1214-277   &	 H{\sevensize II} TI		& 0.026 & -17.59 \\  
Tol 1008-286   &	 H{\sevensize II} E? TI	& 0.014 & -19.31 \\  
Tol 1304-353   &	 H{\sevensize II} 		& 0.014 & -16.74 \\  
Tol 1334-326   &	 H{\sevensize II}	TI	& 0.012 & -17.61 \\  
Tol 1304-386   &	 H{\sevensize II}        	& 0.014 & \\  
UM568          &	 H{\sevensize II}	 	& 0.048 & \\
UM411          &	 H{\sevensize II}		& 0.040 & \\ \hline 
\multicolumn{4}{l}{{\bf G\_UM461} H{\sevensize II} $0.2 \times 0.3$ kpc  $<$M$_B$$>$$=-15.0 \pm 2.7$}  \\ \hline
UM461          & 	 H{\sevensize II} BCD/Irr TII	& 0.003 & -13.53 \\  
UM309          &	 H{\sevensize II} SAB(s)m 	& 0.004 & -19.11 \\  
UM462NE	       &	 H{\sevensize II} pec;BCD  	& 0.004 &	 \\  
UM462SW	       &	 H{\sevensize II} pec;BCD   	& 0.003 & 	 \\  
UM463          &	 H{\sevensize II} 		& 0.004 & -12.62 \\  
UM439          &	 H{\sevensize II} TII		& 0.004 & -15.92 \\  
II Zw 40       &  	 H{\sevensize II} BCD/Irr/merger TI & 0.003 & -14.92 \\  
Tol 1116-325   &	 H{\sevensize II} TII	& 0.002 & -11.52 \\
Tol 1400-411   &	 H{\sevensize II} IB(s)m	& 0.002 & -17.32 \\ \hline
\multicolumn{4}{l}{{\bf G\_Tol1924-416}  H{\sevensize II} blue  $2.5 \times 3.3$ kpc  $<$M$_B$$>$$=-20.0 \pm 0.9$} \\ \hline
Tol 1924-416   &	 H{\sevensize II} pec    	& 0.009 & -19.48 \\
Cam1543+0907$^*$   &	 		& 0.038 & \\
Cam1409+1200$^*$   &	 TI		& 0.056 & \\
Cam0357-3915$^*$   &	 H{\sevensize II} TII	& 0.075 & \\ 
Tol 1247-232   &	 H{\sevensize II}		& 0.048 & -20.95 \\  
Tol 0513-393   &	 H{\sevensize II} TII	& 0.050 & \\  
Tol 1457-262b  &	 H{\sevensize II}		& 0.017 & -19.46 \\ \hline  
\multicolumn{4}{l}{{\bf G\_NGC1487} H{\sevensize II} $0.2 \times 0.3$ kpc  $<$M$_B$$>$$=-17.1 \pm 1.1$} \\ \hline
NGC1487a$^*$   & S pec (merger)			& 0.003 & -18.00 \\  
NGC1487b$^*$       & S pec (merger)	        & 0.003 & -18.00 \\  
NGC1487c$^*$       & S pec (merger)            & 0.003 & -18.00 \\  
Tol 0957-278   &	 H{\sevensize II} multiple system? & 0.005 & -17.13 \\
UM523          &	 H{\sevensize II} Im pec	& 0.003 & -16.2 \\  
UM038          &	 H{\sevensize II} Sc		& 0.004 & -15.52\\  \hline 
\multicolumn{4}{l}{{\bf G\_Mrk710} Starburst $0.2 \times 0.4$ kpc  $<$M$_B$$>$$=-18.5$} \\ \hline
Mrk710         &	 H{\sevensize II} SB(rs)ab	& 0.005 & -18.47 \\  
UM506          &	 H{\sevensize II}		& 0.005 & \\ \hline 
\multicolumn{4}{l}{{\bf G\_Tol1004-296} H{\sevensize II} $0.4 \times 0.6$ kpc  $<$M$_B$$>$$=-17.2 \pm 1.2$} \\ \hline
Tol 1004-296NW &	 H{\sevensize II} TII:	& 0.003 & \\  
Tol 1004-296SE &   	 H{\sevensize II} 		& 0.004 & \\  
Tol 0633-415   &	 H{\sevensize II} TI         & 0.017 & -17.65 \\  
Tol 1324-276   &	 H{\sevensize II} S? TII:	& 0.006 & -17.46 \\  
Tol 1223-359   & 	 H{\sevensize II}		& 0.009 & -15.31 \\  
Fairall 30     &	 H{\sevensize II} SAB(r)0 TII	& 0.004 & \\  
Cam1148-2020B$^*$  &	 H{\sevensize II}		& 0.012 & \\
UM311          &	 H{\sevensize II}		& 0.006 & \\  
UM372          &	 H{\sevensize II}		& 0.005 & \\  
UM238          &	 H{\sevensize II} S TII	& 0.014 & -17.18 \\  
UM382          &	 H{\sevensize II}		& 0.012 & \\  
UM626          &	 H{\sevensize II}		& 0.011 & \\  
UM619          &	 H{\sevensize II} Sm		& 0.015 & -18.49 \\  
UM408          &	 H{\sevensize II} TII	& 0.018 & \\  
UM069W         &	 H{\sevensize II}		& 0.006 & \\  
UM133          &	 H{\sevensize II} TII	& 0.007 & \\  
UM505          & 	 H{\sevensize II}		& 0.004 & \\ \hline 
 \end{tabular}
\end{table}

\setcounter{table}{0} 
\begin{table}
 \caption{(continued)}
 \begin{tabular}{lrcc} \hline
Designation    & Morphology/   & z & M$_{B}$ \\ 
 & Spectral types & & \\ \hline
\multicolumn{4}{l}{{\bf G\_UM448} H{\sevensize II} $1.8 \times 2.1$ kpc} \\ \hline
UM448    & 	 H{\sevensize II} Sb pec TI	& 0.018 & -19.46 \\ \hline
\multicolumn{4}{l}{{\bf G\_Tol0440-381} H{\sevensize II} $1.5 \times 2.7$ kpc  $<$M$_B$$>$$=-18.6 \pm 0.7$} \\ \hline
Tol 0440-381   &	 H{\sevensize II} TI		& 0.041 & \\  
Tol 1025-284   &	 H{\sevensize II} TI:	& 0.032 & -18.66 \\  
Tol 0645-376   &	 H{\sevensize II} TI		& 0.026 & \\  
Tol 1457-262f  &	 H{\sevensize II}		& 0.017 & \\  
Tol 0620-386   &	 H{\sevensize II}		& 0.067 & \\
Cam0951-1733$^*$ &	                & 0.028 & \\
UM653          &	 H{\sevensize II} Sy2	& 0.038 & -17.91 \\  
UM469          &	 H{\sevensize II}		& 0.058 & -18.83 \\  
UM254          &	 H{\sevensize II} Sy2	& 0.044 & -19.23 \\  
UM471          &	 H{\sevensize II}		& 0.035 & -17.00 \\  
UM401          &	 H{\sevensize II}		& 0.032 & \\  
UM031          &	 H{\sevensize II}		& 0.042 & \\  
UM628          &	 H{\sevensize II}		& 0.024 & \\  
UM377          &	 H{\sevensize II}        	& 0.029 & \\  
UM410S         &	 H{\sevensize II} S		& 0.023 & -18.82 \\  
UM410N         &	 H{\sevensize II} S		& 0.023 & -18.82 \\  
UM048          &	 H{\sevensize II} S		& 0.016 & -18.53 \\  
UM494          &	 H{\sevensize II} cIBm	& 0.028 & -19.95 \\  \hline 
\multicolumn{4}{l}{{\bf G\_UM504} H{\sevensize II} $0.4 \times 0.8$ kpc  $<$M$_B$$>$$=-15.9 \pm 0.7$} \\ \hline
UM504          &	 H{\sevensize II}		& 0.007 & -15.08 \\  
UM330          &	 H{\sevensize II}		& 0.017 & \\  
UM077          &	 H{\sevensize II}		& 0.017 & -15.56 \\  
UM080          &	 H{\sevensize II}		& 0.016 & -16.21 \\  
UM051          &	 H{\sevensize II} Sc		& 0.014 & \\  
UM323          &	 H{\sevensize II} BCD	& 0.006 & -15.22 \\  
UM069E         &	 H{\sevensize II}		& 0.006 & \\  
UM069          & 	 H{\sevensize II}		& 0.006 & \\  
UM513          &	 H{\sevensize II}		& 0.012 & -15.88 \\  
UM491          &	 H{\sevensize II}		& 0.006 & -15.29 \\  
UM040S         &	 H{\sevensize II} SB0?	& 0.004 & -15.72 \\  
Cam0945-2219b$^*$ &			& 0.015	& 	\\
Tol 0527-394   &	 H{\sevensize II}		& 0.015 & \\  
Tol 1147-283   &	 H{\sevensize II} N TII	& 0.007 & -16.33 \\  
Tol 1324-276nuc&	 H{\sevensize II} S?	TII:    & 0.006 & -17.46 \\ \hline 
\multicolumn{4}{l}{{\bf G\_Cam0949-2126} H{\sevensize II}/Starb. $2.5 \times 3.8$ kpc  $<$M$_B$$>$$=-19.7 \pm 1.5$} \\ \hline
Cam0949-2126$^*$ &	                & 0.028 &        \\ 
Tol 1457-262a  &	 H{\sevensize II}		& 0.017 & \\  
Tol 0336-407   &         H{\sevensize II} 		& 0.065 & \\  
Fairall 02     &	 H{\sevensize II} SB(r)a	& 0.055 & -21.35 \\  
Cam0840+1201$^*$ &	 TI		& 0.030 & -17.88 \\
UM150          &	 H{\sevensize II} S		& 0.043 & -19.18 \\  
UM307          &	 H{\sevensize II} Sdm:	& 0.023 & -20.51 \\   
UM083          &	 H{\sevensize II}		& 0.040 & \\   
UM358          &	 H{\sevensize II}		& 0.056 & \\ \hline 
\end{tabular}
\end{table}

\setcounter{table}{0} 
\begin{table}
 \caption{(continued)}
 \begin{tabular}{lrcc} \hline
Designation    & Morphology/   & z & M$_{B}$ \\ 
 & Spectral types & & \\ \hline
\multicolumn{4}{l}{{\bf G\_UM71} H{\sevensize II} $1.1 \times 2.1$ kpc  $<$M$_B$$>$$=-17.7 \pm 1.2$} \\ \hline
UM71           & 	 H{\sevensize II}		& 0.040 & -19.52 \\  
UM441          &	 H{\sevensize II}            & 0.044 & -19.09 \\  
UM061          &	 H{\sevensize II} Sc+	& 0.037 & -18.85 \\  
UM047          &	 H{\sevensize II}		& 0.021 & -18.62 \\  
UM065          &	 H{\sevensize II}		& 0.021 & -17.42 \\  
UM017          &	 H{\sevensize II}		& 0.027 & \\  
UM227          &	 H{\sevensize II} Compact	& 0.030 & -18.90 \\  
UM098          &	 H{\sevensize II}		& 0.023 & \\  
UM582          &	 H{\sevensize II}		& 0.058 & \\  
UM110          &	 H{\sevensize II}		& 0.030 & -18.29 \\  
UM595          &	 H{\sevensize II} IR  	& 0.021 & -18.55 \\  
UM219          &	 H{\sevensize II} SBa	& 0.013 & -17.08 \\  
UM354          &	 H{\sevensize II}		& 0.031 & \\  
UM390          &	 H{\sevensize II}		& 0.034 & \\  
UM151          &	 H{\sevensize II} Compact	& 0.016 & -17.03 \\  
UM612          &	 H{\sevensize II}		& 0.015 & \\  
UM055          &	 H{\sevensize II}		& 0.018 & \\  
UM374          &	 H{\sevensize II}		& 0.019 & \\  
UM351          &	 H{\sevensize II}		& 0.026 & -16.91 \\  
UM442          &	 H{\sevensize II}		& 0.027 & -17.68 \\  
UM306          &	 H{\sevensize II}		& 0.016 & -17.26 \\  
UM147          &	 H{\sevensize II}		& 0.032 & \\  
UM618          &	 H{\sevensize II}		& 0.014 & \\  
UM635          &	 H{\sevensize II}		& 0.025 & \\  
UM345          &	 H{\sevensize II}		& 0.019 & \\  
UM455          &	 H{\sevensize II} TII	& 0.012 & -15.55 \\  
UM376          &	 H{\sevensize II}		& 0.039 & \\  
UM507          &	 H{\sevensize II}		& 0.020 & -15.98 \\  
UM369          &	 H{\sevensize II}		& 0.019 & \\  
Fairall 05     &	 H{\sevensize II} Bright Nucleus	& 0.029 & \\  
Tol 0452-416   &	 H{\sevensize II}		& 0.037 & \\
Cam0841-1610$^*$     &	                & 0.028 & \\ 
Cam0954-1852SE$^*$   &	                & 0.014 & \\
Cam0939-2107$^*$    &	                & 0.015 & \\
Cam0952-2028$^*$     &	                & 0.028 & \\
Tol 1345-420   &	 H{\sevensize II} TII	& 0.008 & -16.23 \\ \hline 
\multicolumn{4}{l}{{\bf G\_NGC1510} H{\sevensize II} $0.3 \times 0.4$ kpc} \\ \hline
NGC1510$^*$ &	 SA0 pec H{\sevensize II}	& 0.003 & -16.93	\\ \hline
\multicolumn{4}{l}{{\bf G\_Mrk711} H{\sevensize II}/Starb. $1.4 \times 2.1$ kpc  $<$M$_B$$>$$=-19.8 \pm 0.8$} \\ \hline
Mrk711         &	 H{\sevensize II} Compact SB nuc.	& 0.019 & -18.9 \\  
UM448W 	       &	 H{\sevensize II} Sb pec:	& 0.018 & -19.59 \\  
Cam08-28A$^*$  &	 Compact TI	& 0.053 & -20.93 \\
Cam0942-1929a$^*$  &	                & 0.033 & \\
UM623          &	 H{\sevensize II}		& 0.034 & \\ \hline 
\end{tabular}
\end{table}

\setcounter{table}{0} 
\begin{table}
\begin{minipage}{80mm}
 \caption{(continued)}
 \begin{tabular}{lrcc} \hline
Designation    & Morphology/   & z & M$_{B}$ \\ 
 & Spectral types & & \\ \hline
\multicolumn{4}{l}{{\bf G\_UM140} H{\sevensize II}/Starb. $0.8 \times 1.5$ kpc  $<$M$_B$$>$$=-18.1 \pm 0.9$} \\ \hline
UM140          & 	 H{\sevensize II} Compact	& 0.017 & -19.03 \\  
UM054          &	 H{\sevensize II}		& 0.016 & -18.44 \\  
UM622          &	 H{\sevensize II}		& 0.025 & -19.50 \\  
UM033          &	 H{\sevensize II}		& 0.029 & \\  
UM119          &	 H{\sevensize II}		& 0.021 & \\  
UM395          &	 H{\sevensize II} S		& 0.022 & -18.27 \\  
UM134          &	 H{\sevensize II}		& 0.018 & \\  
UM063          &	 H{\sevensize II} Sb 		& 0.018 & -17.29 \\  
UM064          &	 H{\sevensize II}		& 0.018 & -17.29 \\  
UM079          &	 H{\sevensize II}		& 0.017 & \\  
UM049          &	 H{\sevensize II}		& 0.014 & \\  
Cam0956-2143$^*$     &	                & 0.016 & \\ 
Cam0956-1916$^*$    &	                & 0.012 & \\
UM512          &	 H{\sevensize II}		& 0.015 & -17.32 \\
UM371          &	 H{\sevensize II}		& 0.018 & -17.64 \\ \hline 
\multicolumn{4}{l}{{\bf G\_NGC3089} H{\sevensize II}/Starb. $1.1 \times 2.1$ kpc  $<$M$_B$$>$$=-19.3 \pm 1.0$} \\ \hline
NGC3089$^*$    &	 SAB(rs)b	& 0.009 & -19.57 \\  
Tol 1025-285   &	 H{\sevensize II} N		& 0.031 & -19.30 \\  
UM308          &	 H{\sevensize II} S0: sp		& 0.016 & -18.53 \\  
UM304          &	 H{\sevensize II} Sb		& 0.016 & -19.03 \\  
UM027          &	 H{\sevensize II}		& 0.034 & -19.92 \\  
UM646          &	 H{\sevensize II}		& 0.025 & -18.65 \\  
UM598          &	 H{\sevensize II} SAB(s)b pec  & 0.022 & -21.75 	\\  
UM078          &	 H{\sevensize II}		& 0.040 & -18.88 \\  
UM391          &	 H{\sevensize II} Compact	& 0.021 & -19.22 \\  
UM203          &	 H{\sevensize II}		& 0.040 & -18.02 \\  
UM601          &	 H{\sevensize II}		& 0.028 & \\  \hline
\multicolumn{4}{l}{{\bf G\_UM477} Starburst $0.7 \times 1.4$ kpc  $<$M$_B$$>$$=-20.1 \pm 0.9$} \\ \hline
UM477          &	 H{\sevensize II} SB(r)c	& 0.004 & -19.04 \\  
UM467          &	 H{\sevensize II} (R)SAB0+	& 0.019 & -18.45 \\  
UM274          &	 H{\sevensize II} SA(rs)b pec? & 0.013 & -20.61 \\  
UM418          &	 H{\sevensize II} (R')SAB(s)b pec & 0.026 & -20.79 \\  
UM152E         &	 H{\sevensize II} Sc 		& 0.018 & -19.79 \\  
UM152W         &	 H{\sevensize II} SBm pec:	& 0.018 & -20.38 \\  
UM093          &	 H{\sevensize II} Compact	& 0.032 & -21.08 \\  
NGC4536 nuc$^*$    &	 SAB(rs)bc H{\sevensize II}	& 0.006 & -20.74 \\ \hline 
\multicolumn{4}{l}{{\bf G\_UM103} Seyfert 2 $1.8 \times 3.7$ kpc  $<$M$_B$$>$$=-19.7 \pm 0.4$} \\ \hline
UM103          & 	 Sy2		& 0.045 & -19.28 \\  
NGC2989$^*$    & 	 SAB(s)bc:	& 0.014 & -20.16 \\  
Tol 0514-415   & 	 Sy2		& 0.049 & \\ \hline 
\multicolumn{4}{l}{{\bf G\_NGC4507} Seyfert 2 $1.7 \times 3.5$ kpc  $<$M$_B$$>$$=-19.8 \pm 0.8$} \\ \hline
NGC4507$^*$    &	 SAB(s)ab   Sy2	& 0.012 & -20.49 \\  
NGC3081$^*$        & 	 (R\_1)SAB(r)0/a  Sy2	& 0.008 & -19.67 \\  
UM082          &	 Sy2		& 0.051 & -18.67 \\  
UM363          &	 Sy2 (R)SAB(rs)0+:  	& 0.018 & -20.61 \\  
UM085          &	 Sy2		& 0.041 & -18.78 \\  
UM293          &	 Sy2 Sy1.5		& 0.056 & -20.35 \\  
UM016          &	 Sy2		& 0.058 & -19.83 \\ \hline 
\multicolumn{4}{l}{{\bf G\_NGC3281} Seyfert 2 $1.2 \times 2.4$ kpc  $<$M$_B$$>$$=-19.7 \pm 1.0$} \\ \hline
NGC3281$^*$        &	 SAB(rs+)a   Sy2 & 0.012 & -20.71 \\  
NGC1386$^*$        & 	 SB(s)0+   Sy2	& 0.003 & -18.31 \\  
UM319          &	 Sy2 SB? 	& 0.016 & -19.15 \\  
UM105          &	 Sy2		& 0.030 & -20.21 \\  
Tol 0611-379   &	 Sy2		& 0.038 & \\  
Tol 0544-395   &	 Sy2 (R)AS(r)0/a & 0.025 & -20.10	\\ \hline 
\end{tabular}
\medskip
$^*$ Additional galaxies with respect to Terlevich et al. (1991).
\end{minipage}
\end{table}

\begin{figure*}
 \vspace{20cm}
 \includegraphics{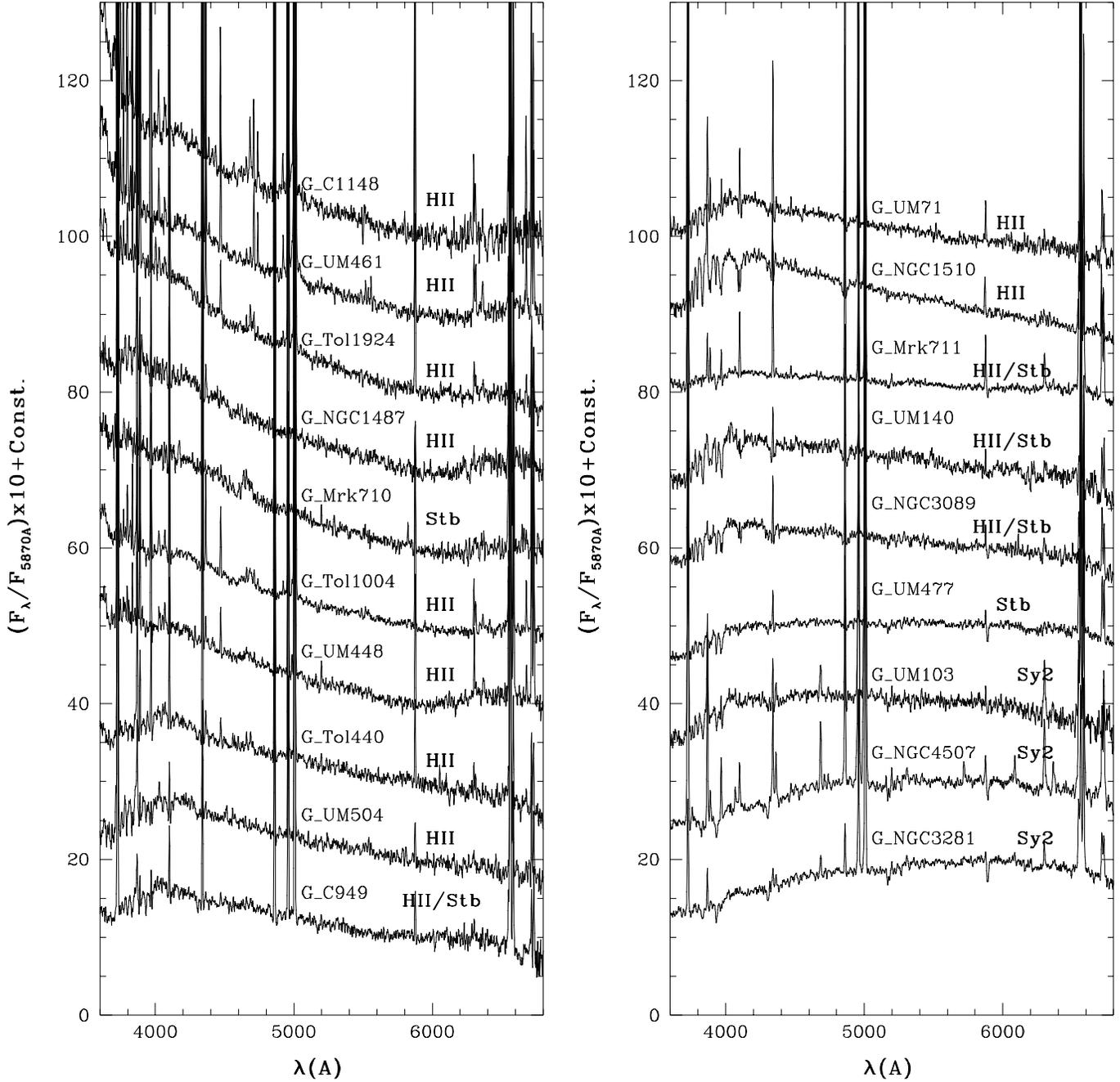}
 \caption{The resulting spectral group templates. The basic emission line nature of the groups as H{\sevensize II} galaxy, Intermediate H{\sevensize II}/Starburst, Starburst and Seyfert 2 are indicated. For some groups designations are abbreviated.}
 \label{Grup}
\end{figure*}

\subsection{The spectral groups}

The resulting spectral group templates are shown in Fig. \ref{Grup}, where the nature of the emission line component is also indicated. They can be basically inferred from information on member galaxies in Table \ref{memb}. However for the sake of clarity what we show in Fig. \ref{Grup} as emission classification is a high precision a {\it posteriori} result (Paper II). It is based on the distribution of emission line spectra of the groups in the Baldwin, Phillips and Terlevich (1981) diagnostic diagram, after subtraction of stellar population models. In addition to the classifications as H{\sevensize II} galaxies, Starbursts and Seyfert 2's, we give Intermediate H{\sevensize II}/Starburst classification for groups near this bordering locus in the diagnostic diagram.

The groups are ordered from bluer to redder continuum distributions in the figure. Note that most H{\sevensize II} groups have indeed very blue continua as expected from stellar populations dominated by very recent star formation, but some H{\sevensize II} groups have a prominent Balmer jump in absorption, suggesting the presence of evolved blue stellar populations. The Intermediate H{\sevensize II}/Starburst groups sit on stellar populations with Balmer jump in absorption and in some cases (e.g. G\_UM140) with an incipient 4000 \AA\ break which is a characteristic of populations older than 1 Gyr. For a detailed discussion of the Balmer jump and 4000 \AA\ break as a function of age and metallicity in single-age stellar populations see Bica et al. (1994). One Starburst group is very blue (G\_Mrk710), while the other (G\_UM477) has a 4000 \AA\ break suggesting an important intermediate age ($1<t<9$ Gyr) and/or old ($t>10$ Gyr) metal-rich components. We can add to the starburst variety of stellar populations very steep spectra rising towards the infrared, such as that of NGC6240 wherein dust absorption considerably affects both recent star formations and the bulge stellar population, as unveiled in the synthesis by Schmitt et al. (1996). Two Seyfert 2 groups are very red, while one (G\_UM103) is somewhat bluer. Finally, note that some very blue H{\sevensize II} groups and one Starburst group (G\_Mrk710) present the wide Wolf-Rayet feature at $\lambda \approx $ 4680 \AA.

\section{Population syntheses}

The star cluster method to synthesize galaxy spectra (Bica, 1988) reduces the number of variables to age and metallicity. The synthesized quantities are W$_\lambda$ of the main absorption spectral features and continuum fluxes (F$_{\lambda}$). These quantities are measured in each spectrum and are compared with combinations of the base elements which in turn have known ages and metallicities. An algorithm generates these combinations in a selected step and compares the synthesized W$_\lambda$ and F$_{\lambda}$ with the observed values. The possible solutions are those that reproduce, within allowed limits, the observed $W_{\lambda}$ and $F_{\lambda}$ of the galaxy spectrum. The results obtained through this method are percentage flux fractions (at a selected wavelength) of the different age and metallicity components.   

The algorithm used is an upgraded version (Schmitt et al., 1996). It is faster than previous versions and also solves for the internal reddening $E(B-V)_{i}$ that affects the stellar population.

The employed W$_{\lambda}$ were those of K Ca{\sevensize II} (corresponding to the spectral window $\lambda\lambda3908-3952$ \AA), CN-band ($\lambda\lambda4150-4214$ \AA), G-band ($\lambda\lambda4284-4318$ \AA) and Mg{\sevensize I} ($\lambda\lambda5156-5196$ \AA). The continuum fluxes were measured at $\lambda=$ 3660 \AA, 4020 \AA, 4510 \AA\ and 6630 \AA, normalized at $\lambda=$ 5870 \AA. Table \ref{data} shows these quantities for the 19 groups.

\begin{table*}
 \centering
 \begin{minipage}{140mm}
 \caption{Absorption line equivalent widths and F$_{\lambda}$ relative continuum fluxes for the spectral groups.}
 \label{data}
 \begin{tabular}{lrrrrrrrr} \hline
Group & W(K Ca{\sevensize II}) & W(CN band) & W(G band) & W(Mg{\sevensize I}) & $\frac{F_{3660}}{F_{5870}}$ & $\frac{F_{4020}}{F_{5870}}$ &
$\frac{F_{4510}}{F_{5870}}$ & $\frac{F_{6630}}{F_{5870}}$ \\ \hline
G\_Cam1148-2020   & -0.2& -0.3& 0.0 & -0.1& 3.33 & 2.27 & 1.72 & 1.09 \\
G\_UM461   & 0.7 & -1.9& -0.7& 0.8 & 3.03 & 2.21 & 1.77 & 1.07 \\ 
G\_Tol1924-416 & 1.0 & -0.7& 0.7 & 1.0 & 2.64 & 2.42 & 1.76 & 0.86 \\
G\_NGC1487 & 1.0 & 0.2 & 0.0 & 0.7 & 2.34 & 2.28 & 1.79 & 1.17 \\
G\_Tol1004-296 & 2.1 & -1.2& 0.1 & 0.0 & 2.23 & 1.97 & 1.62 & 0.94 \\
G\_UM448   & 2.1 & -2.5& -0.8& 2.3 & 2.01 & 1.90 & 1.58 & 1.10 \\
G\_Tol0440-381  & 1.3 & -1.0& 1.4 & 1.1 & 1.58 & 1.78 & 1.43 & 0.69 \\
G\_UM504   & 2.1 & -0.3& 2.3 & 2.5 & 1.40 & 1.70 & 1.48 & 0.79 \\
G\_UM71    & 2.9 & -0.5& 2.3 & 1.9 & 1.10 & 1.53 & 1.34 & 0.74 \\
G\_NGC1510 & 3.0 & 0.1 & 1.4 & 2.2 & 1.06 & 1.72 & 1.58 & 0.79 \\
G\_Cam0949-2126    & 4.0 & 3.3 & 3.7 & 3.4 & 1.26 & 1.66 & 1.42 & 0.75 \\
G\_Mrk711  & 3.7 & -0.8& 1.3 & 2.1 & 1.06 & 1.20 & 1.14 & 0.97 \\
G\_UM140   & 2.2 & 1.1 & 1.9 & 3.7 & 0.86 & 1.35 & 1.31 & 0.80 \\
G\_NGC3089 & 2.5 & 0.1 & 3.0 & 2.8 & 0.82 & 1.30 & 1.19 & 0.81 \\
G\_Mrk710  & 1.5 & -1.6& 0.1 & 1.1 & 2.31 & 2.06 & 1.77 & 0.95 \\
G\_UM477   & 2.6 & -0.5& 2.3 & 3.7 & 0.60 & 0.91 & 1.04 & 0.90 \\
G\_UM103   & 5.3 & 3.3 & 4.1 & 2.6 & 0.59 & 1.07 & 1.11 & 0.72 \\ 
G\_NGC4507 & 13.7& 7.6 & 7.6 & 5.7 & 0.49 & 0.66 & 0.88 & 0.84 \\
G\_NGC3281 & 13.6& 5.6 & 6.7 & 6.5 & 0.35 & 0.56 & 0.77 & 0.83 \\ \hline
\end{tabular}

\medskip
Note. W$_\lambda$ in \AA.
\end{minipage}
\end{table*}

We point out that for the spectral groups within the stellar population range of galaxy spectra discussed in Bica (1988) we employed essentially the same component base, whereas for bluer spectral groups (Sect. 3.2) we adopted a new base with higher time resolution for young components. In the following discussions we will refer to them as Base I and Base II, respectively.

Base I consists of the W$_{\lambda}$ and F$_{\lambda}$ values given by Bica and Alloin (1986b) for single age stellar populations in the age range from H{\sevensize II} regions to globular clusters and metallicities $0.6 \geq [Z/Z_{\odot}] \geq -2.0$. For the H{\sevensize II} region element we used new values: F$_{3660}/$F$_{5870}=3.28$,  F$_{4020}/$F$_{5870}=2.20$, F$_{4510}/$F$_{5870}=1.73$, F$_{6630}/$F$_{5870}=0.74$ which are based on a larger number of observed H{\sevensize II} regions.

Base II consists of the W$_{\lambda}$ and F$_{\lambda}$ values derived from the spectral observations of star clusters of the Large Magellanic Cloud, and derived templates (Santos et al., 1995). These templates have sub-solar metallicity ([Fe/H]$ \approx 0.5$) which is suitable for H{\sevensize II} galaxies, and have ages in the range 2 to 65 Myr. We complemented them with some elements from Base I ($t=100$, 500 Myr and 10 Gyr). Base II is presented in Table \ref{base2}. This new base has a normalization wavelength at $\lambda=4020$ \AA, because the spectra have a spectral range $\lambda3600$ \AA\ to $\lambda5500$ \AA, thus not reaching the previous normalization wavelength $\lambda=5870$ \AA. More details on the implementation of Base II in the synthesis algorithm are given in Raimann (1998).

\begin{table*}
 \centering
 \begin{minipage}{150mm}
 \caption{Base II properties.}
 \label{base2}
 \begin{tabular}{ccrrrrrrrr} \hline
Age (Myr) & Template & W(K Ca{\sevensize II}) & W(CN band) & W(G band) & W(Mg{\sevensize I}) & $\frac{F_{3660}}{F_{4020}}$ & $\frac{F_{3780}}{F_{4020}}$ &  $\frac{F_{4510}}{F_{4020}}$ &  $\frac{F_{5313}}{F_{4020}}$ \\ \hline
2-3   & HIIy.LMC\footnote{Template in Santos et al. (1991).} & 0.00 & 0.00 & 0.00 & 0.00 & 1.60 & 1.40 & 0.80 & 0.5 \\
3-5A  & HIIo.LMC$^a$ & 0.00 & 0.00 & 0.00 & 0.00 & 1.50 & 1.20 & 0.73 & 0.5 \\
3-5B  & YA\_SG.LMC$^a$ & 2.10 & 2.90 & 1.10 & 1.20 & 1.06 & 1.12 & 0.79 & 0.55 \\
6-9   & YB.LMC$^a$   & 0.99 & 0.58 & 1.23 & 1.31 & 0.67 & 0.80 & 0.80 & 0.57 \\
12-40 & YC.LMC$^a$   & 1.87 & 0.48 & 0.42 & 0.15 & 0.67 & 0.94 & 0.75 & 0.53 \\
35-65 & YDE.LMC$^a$  & 2.60 & 0.82 & 0.67 & 0.78 & 0.52 & 0.89 & 0.79 & 0.51 \\
100   & Y3\footnote{Template in Bica (1988).}       & 3.80 & 2.20 & 1.20 & 2.50 & 0.51 & 0.76 & 0.78 & 0.65 \\
500   & Y4$^b$       & 9.20 & 6.10 & 4.40 & 4.50 & 0.50 & 0.66 & 0.91 & 0.83 \\ 
10000   & G1$^b$     & 17.30 & 12.0 & 9.30 & 7.40 & 0.65 & 0.67 & 1.42 & 1.72 \\ \hline 
\end{tabular}
\end{minipage}
\end{table*}

We point out that in all present spectra emission lines are strong and Balmer lines as a rule have both absorption and emission components. We do not use Balmer lines in the syntheses. However metal lines in the blue-visible range like K Ca{\sevensize II} and Mg{\sevensize I} are fiducial age constraints because of dilution effects (Bica, 1988), capable of indicating the presence and amount of intermediate age and/or old components.   

\subsection{Groups synthesized with Base I}

Following Bica (1988) and Schmitt et al. (1996), the first synthesis step is to map out the possible solution space in the plane $t\times[Z/Z_{\odot}]$ which provides an indication of the enrichment level attained by the stellar population components. Dominant components are identified, which helps constrain an evolutionary path in the plane. Subsequently syntheses were perfomed using a selected path. Path 1, Path 2 and Path 3 attain metallicities respectively of $[Z/Z_{\odot}]=0.0$, +0.3 and +0.6 for the intermediate age and young components. In the old bin (10 Gyr) the paths span from $[Z/Z_{\odot}]=-2.0$ up to the respective upper level above. For a visualization of such evolutionary paths in the plane metallicity vs. age see Bica (1988). 

For each synthesis we probe possible internal reddening values affecting the stellar population as a whole. They are in the range $0 \leq E(B-V)_{i} \leq 0.50$ with 0.02 as step. The contribution of each base component (flux fractions at 5870 \AA) varies from 0 to 100 per cent, initially with a step of 10 per cent. In this stage we test allowed limits of each feature equivalent width and continuum point.

Finally, we run the synthesis algorithm with step of 5 per cent for the base components. For the present galaxy spectra only paths 1 and 2 gave solutions. For each reddening value Path 1 (12 components) tests about 80 million combinations and by considering all reddening values it tests 2 billion combinations. Typically we find 200000 solutions, which implies that 1 out of 10000 combinations is a solution. For Path 2 (13 components) five billion combinations are tested and typically 1 million solutions satisfy the allowed limits, implying a 1:5000 solution to combination ratio.

The synthesis results for the groups are presented in Table \ref{SynRes1} in the form of percentage flux contribution from each age and metallicity component to the light at $\lambda=$ 5870 \AA. These flux fractions are averages of all solutions found. As in previous syntheses (e.g. Bica, 1988 and Bonatto et al., 1998, 1999) these values vary within 10 per cent among most solutions. The flux fractions are to be taken as probabilities of the presence of the respective component and values smaller than 5 per cent must be taken with care. The last column shows the internal reddening $E(B-V)_{i}$ affecting the stellar population obtained from the synthesis. The RMS of reddening values among the solutions is typically $\sigma(E(B-V)_{i})=0.03$.

The synthesis residuals  ($\delta=Observed-Synthesized$) for $W_\lambda$ and continuum points are in Table \ref{Resid1}. The residuals are usually $\leq$  2 \AA\ for the W$_\lambda$ and $\leq$ 0.05 for the continua.

\begin{figure}
 \vspace{8cm}
\includegraphics{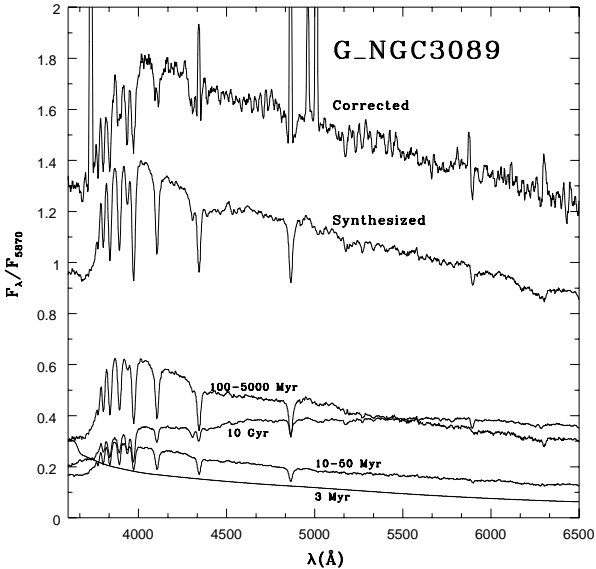}
\caption{Spectrum of the group G\_NGC3089 corrected for reddening $E(B-V)_{i}$, the synthesized spectrum, and synthetic component spectra grouped by age ranges. The reddening corrected spectrum has been shifted by a constant C=+0.4 for clarity.}
 \label{Resul1}
\end{figure}

An illustration of synthesis with Base I is presented in Fig. \ref{Resul1}, which shows the dereddened spectrum of the Intermediate H{\sevensize II}/Starburst group G\_NGC3089 (using the $E(B-V)_{i}$ from Table \ref{SynRes1}), together with the synthesized spectrum constructed by the sum of the base spectra according to the proportions obtained from the synthesis (Table \ref{SynRes1}). We also provide component spectra grouped by age ranges, shown in relative proportions according to the synthesis. 

We point out that the synthetic spectra are built with components which are available template spectra, associated as much as possible with grid elements, which are in turn those effectively used in the synthesis calculations. Note that both in the grid and in the template spectrum, the H{\sevensize II} region element (3 Myr) is a featureless continuum derived from an average of real H{\sevensize II} regions. This featureless continuum is shown as a component in Fig. \ref{Resul1}. Comparing the synthesized spectrum with the galaxy group spectrum (Fig. \ref{Resul1}) the importance of the underlying stellar population absorptions to the study of the emission lines is clear. It can be seen the effect of Na{\sevensize I}$\lambda$5893 \AA\ absorption at the edge of He{\sevensize I}$\lambda$5876 \AA\ in emission. The lines H$\beta \lambda4861$ \AA\ and H$\gamma \lambda4340$ \AA\ appear in emission despite the important underlying absorption. Finally, H$\delta \lambda$4102 \AA, H$\epsilon \lambda$3970 \AA\ and H$\zeta \lambda$3889 \AA\ are seen in absorption, but the synthesis indicates that they must be partially filled by their emission components. On the other hand H$\epsilon$ has a contribution of [Ne{\sevensize III}]$\lambda$3968 \AA\ and an absorption contribution of H Ca{\sevensize II}, since K Ca{\sevensize II}$\lambda$3933 \AA\ is present. In Paper II we will study the pure emission line spectra after subtraction of the present absorption model spectra.

\tabcolsep 3pt
\begin{table*}
 \centering
 \begin{minipage}{160mm}
 \caption{Groups synthesized with Base I: percentage contribution to F$_{\lambda5870}$.}
 \label{SynRes1}
 \begin{tabular}{llrrrrrrrrrrc} \hline
Group & Path & 3 Myr & 10 Myr  &  50 Myr & 100 Myr & 500 Myr & 1 Gyr &  5 Gyr & 10 Gyr &  10 Gyr & 10 Gyr & $E(B-V)_{i}$ \\
  &   &  & & & & & & & $+0.3$\footnote{Metallicity $[Z/Z_{\odot}]$} &  $0.0^{a}$ & $<0.0^{a}$       &  \\ \hline
G\_UM504   & 2 & 24&  2 & 14 & 41 &  5  & 2  & 1  &  1 &  1 &  9  & 0.02   \\ 
G\_UM71    & 1 & 15&  0 &  0 & 51 & 12  & 1  & 1  &     &  1 & 19  & 0.02   \\
G\_NGC1510 & 1 & 9 &  1 & 23 & 59 &  2  & 1  & 1  &   &  0 &  4  & 0.01   \\
G\_Cam0949-2126    & 2 & 18&  2 & 16 & 38 &  9  & 3  & 2  &  1 &  2 &  9  & 0.01   \\
G\_Mrk711  & 1 & 23& 13 & 16 & 13 &  4  & 2  & 2  &     &  2 & 25  & 0.18   \\
G\_UM140   & 1 & 10&  0 &  1 & 57 &  4  & 1  & 2  &     &  7 & 18  & 0.00   \\
G\_NGC3089 & 1 & 8 &  1 & 14 & 35 &  0  & 1  & 2  &     &  6 & 33  & 0.06   \\
G\_UM477   & 1 & 11&  1 &  4 & 56 &  3  & 2  & 2  &     &  8 & 13  & 0.30   \\
G\_UM103   & 2 & 0 &  0 &  1 & 28 & 12  & 1  & 2  &  4 &  5 & 47  & 0.04   \\
G\_NGC4507 & 2 & 1 &  1 &  1 &  1 &  2  & 6  & 9  & 19 & 18 & 42  & 0.05   \\
G\_NGC3281 & 2 & 0 &  0 &  0 &  0 &  0  & 0  & 0  & 24 & 23 & 53  & 0.00   \\ \hline
\end{tabular}
\end{minipage}
\end{table*}
\tabcolsep 6pt

\begin{table*}
 \centering
 \begin{minipage}{150mm}
 \caption{Groups synthesized with Base I: residuals.}
 \label{Resid1}
 \begin{tabular}{lrrrrrrrr} \hline
Group &  $\delta_{W(K CaII)}$ & $\delta_{W(CN band)}$ & $\delta_{W(G band)}$ & $\delta_{W(MgI)}$ & $\delta_{\frac{F_{3660}}{F_{5870}}}$ & $\delta_{\frac{F_{4020}}{F_{5870}}}$ &  $\delta_{\frac{F_{4510}}{F_{5870}}}$ &  $\delta_{\frac{F_{6630}}{F_{5870}}}$ \\ \hline         
G\_UM504   & -1.5 & -2.6 & 0.8 & 0.1  & -0.01 & -0.01 & 0.03 & -0.04 \\ 
G\_UM71    & -1.1 & -2.7 & 0.6 & -0.4 & -0.08 & -0.09 & -0.06 & -0.10 \\ 
G\_NGC1510 & -0.6 & -2.0 & 0.1 & -0.2 & -0.01 & -0.07 & 0.09 & -0.04 \\
G\_Cam0949-2126    & -0.3 & 0.4  & 1.9 & 0.6  & 0.00  & 0.01  & 0.01 & -0.09 \\
G\_Mrk711  & -0.7 & -2.7 & -0.9& 0.1  & 0.00  & 0.01  & 0.03  & 0.04 \\
G\_UM140   & -2.3 & -1.7 & -0.1 & 0.8 & 0.00  & -0.05 & 0.05 & -0.08 \\ 
G\_NGC3089 & -2.1 & -2.5 & 0.8  & 0.1 & -0.01 & 0.00  & -0.01 & -0.09 \\
G\_UM477   & -1.7 & -3.1 & 0.3 & 0.9  & -0.05 & -0.14 & 0.05 & -0.07 \\ 
G\_UM103   & -1.6 & -0.6 & 0.7 & -0.8 & -0.05 & -0.02 & 0.03 & -0.19 \\ 
G\_NGC4507 &  1.3 & -0.6 & 0.7 & -0.1 & 0.02  & -0.03 & 0.03 & -0.14 \\
G\_NGC3281 & -0.2 & -2.9 & -0.9 & 0.6 & -0.08 & -0.05 & -0.05 & -0.15 \\ \hline
\end{tabular}
\end{minipage}
\end{table*}

\subsubsection{Results}

The H{\sevensize II} groups G\_UM71 and G\_UM504, the Intermediate H{\sevensize II}/Starburst groups G\_Cam0949-2126, G\_UM140, G\_Mrk711 and G\_NGC3089, together with the Starburst one G\_UM477 have an old stellar population contributing with about 10 to 40 per cent of the total flux at $\lambda=$ 5870 \AA. All these groups despite the strong emission lines are evolved galaxies dating back from the early universe. These groups have a flux excess at $t \approx 100$ Myr indicating a burst of star formation, except G\_Mrk711 where flux fractions are more evenly distributed among age components.

G\_NGC1510 (H{\sevensize II}) has a marginally detected old component (less than 5 per cent of the total flux at $\lambda=$ 5870 \AA). It presents a burst around 100 Myr, contributing with almost 60 per cent of the total flux at $\lambda=$ 5870 \AA.

Seyfert 2 galaxies may contain some contribution of a featureless continuum in the near-UV. Recent studies of the stellar population in Seyfert 2's, some in common with the present study, with better spatial resolution indicate, however, that the featureless continuum (or combined to H{\sevensize II} region) contribution amounts to less than 10 per cent at $\approx$ 3700 \AA\ (Cid Fernandes et al., 1998; Storchi-Bergmann et al., 1998; Schmitt et al., 1999). In the present cases such contribution, if any, cannot be distinguished from that of H{\sevensize II} regions.    

The Seyfert 2 groups G\_NGC3281, G\_NGC4507 and G\_UM103 have a dominant old stellar population (Table \ref{SynRes1}). They include most galaxies with late-type morphological type, in agreement with the strong old component which is characteristic of bulges. However group G\_UM103 has an additional burst of star formation with age $t \approx 100$ Myr contributing with $\approx$ 27 per cent of the total flux at $\lambda=$ 5870 \AA. This confirms previous results that Seyfert 2 nuclei can sit either on typical old bulge populations or include in the surroundings star-forming regions or evolved bursts of star formation. Indeed Storchi-Bergmann et al. (1990) found this effect in scales of 2--5 kpc around the nucleus, while Cid Fernandes et al. (1998),  Storchi-Bergmann et al. (1998) and Schmitt et al. (1999) found them in spatially resolved regions much closer to the nucleus.    

We find a low internal reddening (column 13 of Table \ref{SynRes1}) for the Seyfert 2 groups. The same occurs for the other groups, except the Intermediate H{\sevensize II}/Starburst group G\_Mrk711 and the Starburst group G\_UM477 which have important reddening $E(B-V)_{i}$ = 0.18 and 0.30, respectively.

In column 2 of Table \ref{SynRes1} we indicate the path which best describes the chemical enrichment. Most groups (especially H{\sevensize II}'s) are described by a solar path (Path 1), but some (especially Seyfert 2's) follow the slightly more metal-rich Path 2.

Finally, it is interesting that on the average the contribution of old populations increases from the H{\sevensize II} groups synthesizable with Base I to the Intermediate H{\sevensize II}/Starburst groups. This suggests that we are dealing with a sequence of increasingly evolved galaxies.

\subsection{Groups synthesized with Base II}

The six elements coming from the spectral templates in Santos et al. (1995) are HIIy.LMC, HIIo.LMC, YA\_SG.LMC, YB.LMC, YC.LMC and YDE.LMC. Note that the first (younger) and second elements are H{\sevensize II} regions with underlying star clusters in two different evolutionary stages and we treate them as featureless continua. The YA\_SG.LMC template is a frequent spectral type among LMC clusters. It has essentially the same age range as the HIIo.LMC template, but it differs from the latter template by having an important contribution of hot supergiants and is basically emission free, possibly as a result of winds (Santos et al., 1995). The three remaining templates are the subsequent evolutionary stages of star clusters spanning the range 6 to 65 Myr. The template W$_\lambda$ and continuum points are in Table \ref{base2}.

Note that the present galaxy groups are extremely blue and not much flux from red stellar population components is expected. So we simply employed a red element of 10 Gyr which could in fact represent any intermediate age ($t>1$ Gyr) or old component. The continuum points renormalized to $\lambda$4020 \AA\ for the bluer groups, which are subject of syntheses with Base II are given in Table \ref{Cont2}. We recall that the respective W$_\lambda$ are in Table \ref{data}.

\begin{table}
 \centering
 \caption{F$_{\lambda}$ continuum fluxes relative to F$_{4020}$.}
 \label{Cont2}
 \begin{tabular}{lrrrr} \hline
Group &  $\frac{F_{3660}}{F_{4020}}$ & $\frac{F_{3780}}{F_{4020}}$ &  $\frac{F_{4510}}{F_{4020}}$ &  $\frac{F_{5313}}{F_{4020}}$ \\ \hline          
G\_Cam1148-2020   & 1.43 & 1.14 & 0.75 & 0.54   \\
G\_UM461   & 1.38 & 1.14 & 0.80 & 0.58  \\
G\_Tol1924-416 & 1.11 & 1.09 & 0.72 & 0.51 \\ 
G\_NGC1487 & 1.03 & 1.09 & 0.80 & 0.56 \\ 
G\_Tol1004-296 & 1.14 & 1.03 & 0.80 & 0.57 \\ 
G\_UM448   & 1.07 & 1.12 & 0.82 & 0.62  \\
G\_Tol0440-381  & 0.85 & 0.87 & 0.79 & 0.63 \\ 
G\_Mrk710  & 1.08 & 1.11 & 0.84 & 0.55  \\  \hline
\end{tabular}
\end{table}

As for Base I we test the possibility of internal reddening $0 \leq E(B-V)_{i} \leq 0.50$ with 0.02 as step. Base II has nine components which total about 75 million combinations with a 5  per cent step for the flux fractions. Typically we obtain 4000 solutions which imply a solution to combination ratio of 1:18750. 

\begin{figure}
 \vspace{8.0cm}
\includegraphics{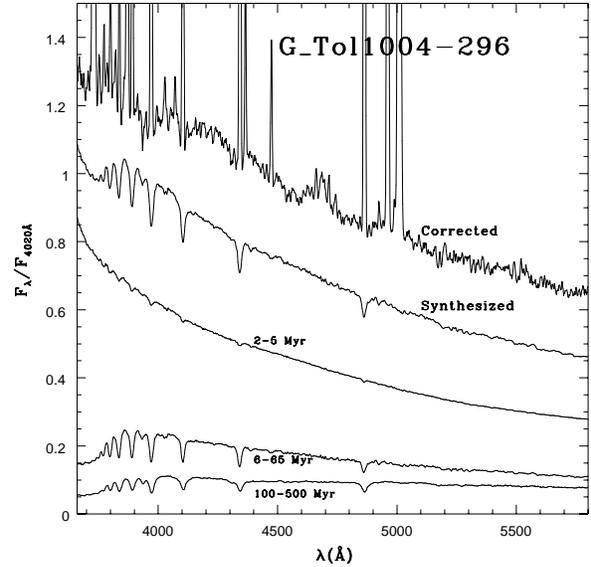}
 \caption{Spectrum of the group G\_Tol1004-296 corrected for reddening $E(B-V)_{i}$, the synthesized spectrum, and synthetic component spectra grouped by age ranges. The reddening corrected spectrum has been shifted by constant C=+0.2 for clarity.}
 \label{Resul2}
\end{figure}

The synthesis results for these groups are presented in Table \ref{SynRes2} and the residuals in Table \ref{Resid2}. As an example, Fig. \ref{Resul2} shows the dereddened group G\_Tol1004-296 according to the synthesis, the corresponding synthesis model and component spectra grouped by age ranges. The underlying stellar population affects less the emission line spectrum as compared to the example given in Fig. \ref{Resul1}.

It is worth remarking that the spectral boundary between the bluer and redder galaxy groups occurs for G\_Tol0440-381 (synthesized with Base II) and G\_UM504 (with Base I). Although there is no apparent color break at this boundary (Fig. \ref{Grup}) the spectral properties of G\_Tol0440-381 cannot be described by combinations of elements in Base I. A comparison of the syntheses results between G\_Tol0440-381 (Table \ref{SynRes2}) and G\_UM504 (Table \ref{SynRes1}) considering equivalent age ranges can be used to test the continuity of the syntheses for the two bases. The populations related to emission contribute respectively with 33 and 24 per cent while the blue star cluster components (6 $\leq$ t $\leq$ 500 Myr) have respectively 66 and 62 per cent, thus similar results. The old populations contribute respectively with 1 and 11 per cent, which is at the limit of the significance level (section 3.1). Indeed the W$_{\lambda}$'s and F$_{\lambda}$'s measurements (Table \ref{data}) show significant differences between these two groups, in agreement with a population of slightly later type for G\_UM504.

\tabcolsep 3pt
\begin{table*}
 \centering
 \begin{minipage}{160mm}
 \caption{Groups synthesized with Base II: percentage contribution to F$_{\lambda4020}$.}
 \label{SynRes2}
 \begin{tabular}{lrrrrrrrrrrrrc} \hline
Group &  2-3 Myr & 3-5A Myr & 3-5B Myr &  6-9 Myr & 12-40 Myr & 35-65 Myr & 100 Myr &  500 Myr & 10 Gyr & $E(B-V)_{i}$ \\ \hline
G\_Cam1148-2020    & 12 & 77 & 3  & 2  & 1  & 1  & 1  & 1 & 2 & 0.06 \\
G\_UM461           & 9  & 79 & 4  & 2  & 2  & 1  & 2  & 1 & 0 & 0.13 \\
G\_Tol1924-416     & 15 & 34 & 12 & 4  & 16 & 10 & 3  & 1 & 5 & 0.01 \\
G\_NGC1487         & 31 & 9  & 12 & 9  & 18 & 15 & 5  & 1 & 0 & 0.06 \\
G\_Tol1004-296     & 11 & 46 & 8  & 15 & 6  & 6  & 5  & 3 & 0 & 0.07 \\
G\_UM448           & 38 & 14 & 2  & 13 & 15 & 14 & 4  & 0 & 0 & 0.16 \\
G\_Tol440-381      & 3  & 24 & 6  & 24 & 15 & 5  & 17 & 5 & 1 & 0.10 \\
G\_Mrk710          & 40 & 6  & 11 & 10 & 9  & 17 & 4  & 3 & 0 & 0.05 \\ \hline
\end{tabular}
\end{minipage}
\end{table*}
\tabcolsep 6pt

\begin{table*}
 \centering
 \begin{minipage}{150mm} 
 \caption{Groups synthesized with Base II: residuals.}
 \label{Resid2}
 \begin{tabular}{lrrrrrrrr} \hline
Group & $\delta_{W(K Ca II)}$ & $\delta_{W(CN band)}$ & $\delta_{W(G band)}$ & $\delta_{W(MgI)}$ & $\delta_{\frac{F_{3660}}{F_{4020}}}$ & $\delta_{\frac{F_{3780}}{F_{4020}}}$ &  $\delta_{\frac{F_{4510}}{F_{4020}}}$ &  $\delta_{\frac{F_{5313}}{F_{4020}}}$ \\ \hline         
G\_Cam1148-2020   & -0.7 & -0.8 & -0.3 & -0.7 &  0.04 & -0.04 & -0.04 & -0.01 \\
G\_UM461   &  0.4 & -2.1 & -0.8 &  0.5 &  0.02 &  0.01 & -0.02 &  0.02 \\ 
G\_Tol1924-416 & -0.7 & -2.0 & -0.3 & -0.4 & -0.02 &  0.04 & -0.03 & -0.04 \\ 
G\_NGC1487 & -0.3 & -0.6 & -0.6 &  0.0 & -0.04 &  0.05 &  0.00 &  0.01 \\ 
G\_Tol1004-296 &  1.1 & -1.9 & -0.5 & -0.7 & -0.01 &  0.03 &  0.00 &  0.00 \\ 
G\_UM448   &  1.2 & -2.9 & -1.2 &  1.8 & -0.06 &  0.07 & -0.01 &  0.02 \\ 
G\_Tol0440-381  & -0.7 & -2.3 &  0.4 & -0.4 &  0.00 & -0.03 & -0.04 &  0.02 \\
G\_Mrk710  &  0.2 & -2.5 & -0.5 &  0.4 & -0.04 &  0.05 &  0.03 &  0.00 \\ 
 \hline
\end{tabular}
\end{minipage}
\end{table*}

\subsubsection{Results}

Most blue groups are dominated by components younger than 5 Myr, which is consistent with the fact that they have very strong emission line spectra (Fig. \ref{Grup}). But note that these groups are not single aged, because important flux fractions often occur to ages as old as 100 Myr. Finally, we point out that no red stellar population is allowed by the synthesis for most of these groups. Three groups (especially G\_Tol1924-416) show a marginal detection but at such levels the red component could represent any stellar population in the range $1\leq t \leq 10$ Gyr. The internal reddening affecting the stellar populations $(E(B-V)_{i})$ is low, with the higher value $E(B-V)_{i}=$0.16 for to G\_UM448.

\subsection{Wolf-Rayet feature}

\begin{figure}
\vspace{8.0cm}
\includegraphics{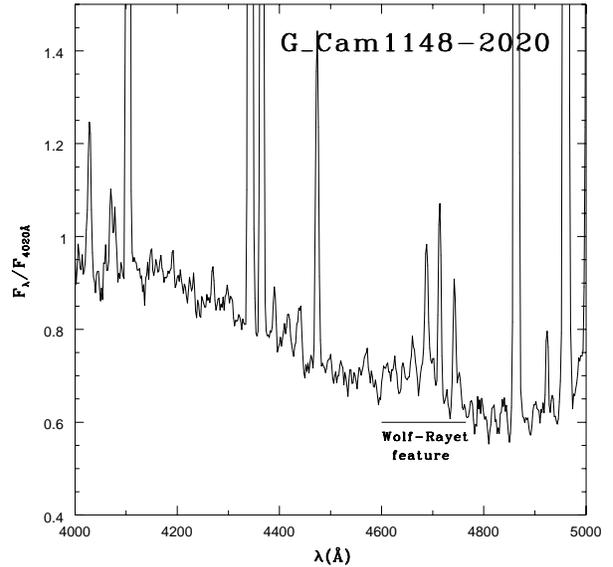}
\caption{Blowup of the Wolf-Rayet feature spectral region for the group G\_Cam1148-2020.}
\label{wr}
\end{figure}

The broad $\lambda\lambda$4650-4690 \AA\ feature, arising in the atmospheres of Wolf-Rayet stars, is present in most of the bluer sample spectra (Fig. \ref{memb}): the H{\sevensize II} groups G\_Cam1148-2020, G\_Tol1004-296,  G\_UM448, G\_Tol1924-416, G\_UM461 and G\_Tol0440-381, and the Starburst group G\_Mrk710. A blowup of the spectral region is shown for G\_Cam1148-2020 in Fig. \ref{wr}. Wolf-Rayet stars are highly evolved products of the evolution of massive stars ($M>35 M_{\odot}$ -- see e.g. Conti et al., 1983 and Humphreys et al., 1985). Because of the limited age range in which they arise, Wolf-Rayet stars can be used as an age indicator in integrated spectra of star clusters (Santos Jr. et al., 1995 and references therein). Schaerer et al. (1996) studied Wolf-Rayet galaxies and used the Wolf-Rayet feature as indicator of star formation with ages younger than 10 Myr.

The population syntheses for the groups with Wolf-Rayet features (Table \ref{SynRes2}) show that the contribution of components younger than 10 Myr varies from 30 to 90 per cent of the total flux at $\lambda=$ 4020 \AA, which explains why the Wolf-Rayet feature is so strong in their spectra.

\subsection{Mass fractions}

From the population synthesis analyses above it is clear that the young components dominate the optical flux in the bluer galaxies. It is also important to compute how much they represent in terms of  mass fractions. For such purposes we employed a flux-mass transformation method (Bica et al., 1988). This method uses different mass to V-light ratios (M/L$_v$) related to each age component. It also takes into account metallicity effects among old star clusters. We show in Tables \ref{Mass1} and \ref{Mass2} mass distributions respectively for the groups synthesized with Base I and Base II. 

In the groups synthesized with Base I (Table \ref{Mass1}), the oldest component (10 Gyr) is dominant in mass ($>$ 60 per cent in all cases). Although the corresponding flux fraction may be as low as a 4 per cent (G\_NGC1510), the mass fraction would be important because the M/L$_v$ ratio is extremely low for young populations. As pointed out in Section 3, uncertainties are important for flux fractions $\leq$ 5 per cent, and they propagate to mass fractions.  

However the scenario is much more favourable for flux and mass fractions related to young components. The mass stocked in the 100 Myr component is important in several spectra: in the case of G\_NGC1510 it is 28 per cent which indicates a strong burst of star formation, and $\approx$ 10 per cent for the groups G\_Cam0949-2126, G\_UM504, G\_UM71, G\_UM477 and G\_UM140.

In the groups synthesized with Base II (Table \ref{Mass2}) we show mass fractions relative to the total mass stocked in the blue stellar components ($t \leq 500$ Myr), for the sake of comparisons among all these groups. In three groups (G\_Cam1148-2020, G\_Tol1924-416 and G\_Tol0440-381) an older population has been marginally detected in flux fraction (Table \ref{SynRes2}). In terms of mass, for the reasons explained in the previous paragraph, the fractions stocked in the older populations would of course be dominant ($\approx$ 93, 97 and 76 per cent, respectively).

The groups G\_Cam1148-2020 and G\_UM461 appear to have a strong burst of star formation at $t\approx$ 4 Myr, while G\_Tol1924-416 has a considerable mass fraction stocked in the same age. Groups G\_Tol0440-381, G\_NGC1487 and G\_UM448 present an important mass excess at $t\approx$ 100 Myr, while the H{\sevensize II} group G\_Tol1004-296 and the Starburst group G\_Mrk710 have a rather uniform mass distribution as a function of age.

\begin{table*}
 \centering
 \begin{minipage}{160mm}
 \caption{Groups synthesized with Base I: percentage mass fractions.}
 \label{Mass1}
 \begin{tabular}{llrrrrrrrrrr} \hline
Group & Path & 3 Myr & 10 Myr  &  50 Myr & 100 Myr & 500 Myr & 1 Gyr &  5 Gyr & 10 Gyr &  10 Gyr & 10 Gyr \\
  &   &  & & & & & & & $+0.3$ &  $0.0$ & $<0.0$ \\ \hline
G\_UM504   & 2 & 1 &  0 &  1 & 10 &  1  & 2  & 3  &  7 &  9 & 66   \\ 
G\_UM71    & 1 & 0 &  0 &  0 &  8 &  2  & 1  & 1  &     &  3 & 85   \\
G\_NGC1510 & 1 & 1 &  0 &  2 & 28 &  1  & 2  & 4  &     &  6 & 56   \\
G\_Cam0949-2126    & 2 & 0 &  0 &  1 &  9 &  3  & 3  & 6  &  8 &  9 & 61   \\
G\_Mrk711  & 1 & 0 &  0 &  0 &  2 &  1  & 1  & 2  &     &  8 & 86   \\
G\_UM140   & 1 & 0 &  0 &  0 &  7 &  1  & 1  & 3  &     & 24 & 65   \\
G\_NGC3089 & 1 & 0 &  0 &  0 &  3 &  0  & 0  & 2  &     & 13 & 82   \\
G\_UM477   & 1 & 0 &  0 &  0 &  8 &  1  & 1  & 4  &     & 32 & 54   \\
G\_UM103   & 2 & 0 &  0 &  0 &  2 &  1  & 0  & 1  &  6 &  6 & 84   \\
G\_NGC4507 & 2 & 0 &  0 &  0 &  0 &  0  & 1  & 4  & 22 & 21 & 52   \\
G\_NGC3281 & 2 & 0 &  0 &  0 &  0 &  0  & 0  & 0  & 24 & 22 & 54   \\ \hline
\end{tabular}
\end{minipage}
\end{table*}

\begin{table*}
 \centering
 \begin{minipage}{150mm} 
 \caption{Groups synthesized with Base II: percentage mass fractions for the stellar generations with $t \leq 500$ Myr.}
 \label{Mass2}
 \begin{tabular}{lrrrrrrrr} \hline
Group &  2-3 Myr & 3-5A Myr &3-5B Myr &  6-9 Myr & 12-40 Myr & 35-65 Myr & 100 Myr &  500 Myr \\ \hline
G\_Cam1148-2020\footnote{Groups where an old population component has been marginally detected (see text).} &  8     & 57 & 3    &1     &2   &1    &10   &18  \\
G\_UM461           & 6      &57 & 3    &1     &2   &2    &14  &15   \\
G\_Tol1924-416$^a$ & 9      &23 & 8    &1     &20  &12   &23  &4  \\
G\_NGC1487         & 15     &5  & 7    &1     &19  &15   &32  &6    \\
G\_Tol1004-296     & 5      &22  & 5   &2     &5   &5    &27  &29   \\
G\_UM448           & 22     &9  & 1    &2     &19  &16   &31  &0    \\
G\_Tol0440-381$^a$ & 1      & 6 & 2    &2     &8   &3    &53  &25    \\
G\_Mrk710           & 18    & 3 & 5    &1     &8   &15   &23  &27    \\ \hline
\end{tabular}
\end{minipage}
\end{table*}

\section{Concluding remarks}

The population syntheses indicate that H{\sevensize II} galaxies in the local universe as a rule are not forming their first generation of stellar population. The diversity of histories of star formation within H{\sevensize II} galaxy groups shows that they are typically age-composite stellar systems, presenting important contributions from generations up to as old as 500 Myr. These galaxies may still be evolved objects in a cosmological sense if old components at low levels could be detected. 

We find evidence of an old underlying stellar population in the H{\sevensize II} galaxy groups G\_UM504 and G\_UM71. These two groups are probably genuine evolved galaxies.

Finally, we point out that H{\sevensize II} galaxies with stellar populations up to 500 Myr may have had as many as 100 stellar generations, assuming a constant star formation rate and that a typical stellar generation lasts 5 Myr. This result is basically consistent with the gas enrichment, because H{\sevensize II} galaxies, although metal deficient, are typically not extremely so, presenting a range of metallicities (Terlevich et al., 1991 and Paper II). Indeed the metals seen in H{\sevensize II} galaxies must come from previous generations of stars since mixing times of the metals ejected by the current or a few previous generations are far too long to be seen now.

\section*{Acknowledgments}
T.S.B., E.B. and H.S. (during part of this work) acknowledge support from the Brazilian Institution CNPq, and D.R. from CAPES. We thank Iranderly F. de Fernandes (as CNPq undergraduate fellow) for work related to this project. We thank the referee Dr. A. Pickles for valuable remarks.

\label{lastpage}

\end{document}